\documentclass[conference]{IEEEtran}
\IEEEoverridecommandlockouts
\usepackage{cite}
\usepackage{amsmath,amssymb,amsfonts}
\usepackage{algorithmic}
\usepackage{graphicx}
\usepackage{textcomp}
\usepackage{amsthm}
\usepackage{xcolor}
\usepackage{bbding}
\usepackage{makecell}
\usepackage{algorithm}  
\usepackage{color}
\usepackage{multirow}
\usepackage{bm}
\usepackage{array}
\usepackage{booktabs}
\usepackage{tabularx,booktabs}
\usepackage{multicol}
\usepackage{cleveref}
\usepackage{balance}
\usepackage{lastpage}
\usepackage{tabulary}
\usepackage{subfigure}
\usepackage{etoolbox}
\usepackage{bbm}
\usepackage{enumitem}
\usepackage{mathrsfs}
\usepackage{multicol}
\usepackage{amsfonts,amssymb}
\usepackage{ulem}
\usepackage{cancel}
\usepackage{setspace}
\usepackage{stackengine}
\usepackage{threeparttable}
\usepackage{tcolorbox}
\usepackage{fancyhdr}
\usepackage[justification=centering]{caption}

\usepackage{amsthm}
\theoremstyle{definition}

\usepackage{nomencl}
\makenomenclature

\def\BibTeX{{\rm B\kern-.05em{\sc i\kern-.025em b}\kern-.08em
    T\kern-.1667em\lower.7ex\hbox{E}\kern-.125emX}}

\newcommand{\blue}[1]{\textcolor{black}{#1}}

\begin{document}

\bstctlcite{IEEEexample:BSTcontrol}

\title{Large Language Model (LLM)-enabled In-context Learning for Wireless Network Optimization: A Case Study of Power Control 
\thanks{Hao Zhou, Chengming Hu, Dun Yuan, Ye Yuan, and Xue Liu are with the School of Computer Science, McGill University, Montreal, QC H3A 0E9, Canada. (mails:{hao.zhou4, chengming.hu, dun.yuan, ye.yuan3}@mail.mcgill.ca, xueliu@cs.mcgill.ca). Di Wu is with the School of Electrical and Computer Engineering, McGill University, Montreal, QC H3A 0E9, Canada. (email: di.wu5@mcgill.ca).
Jianzhong (Charlie) Zhang is with Samsung Research America, Plano, Texas, TX 75023, USA. (email: jianzhong.z@samsung.com).}
}

\author{\IEEEauthorblockN{ Hao Zhou, Chengming Hu, Dun Yuan, Ye Yuan, Di Wu, \\ Xue Liu, \IEEEmembership{Fellow, IEEE}, and Jianzhong (Charlie) Zhang, \IEEEmembership{Fellow, IEEE}. }}

\maketitle

\thispagestyle{fancy}            
\chead{The latest version of this work has been accepted by ICML 2025 Workshop on ML4Wireless, and the revised title is "\textit{Prompting Wireless Networks: Reinforced In-Context Learning for Power Control".}} 

\renewcommand{\headrulewidth}{1pt}      
\pagestyle{plain}

\begin{abstract}
Large language model (LLM) has recently been considered a promising technique for many fields. This work explores LLM-based wireless network optimization via in-context learning.
To showcase the potential of LLM technologies, we consider the base station (BS) power control as a case study, a fundamental but crucial technique that is widely investigated in wireless networks.    
Different from existing machine learning (ML) methods, our proposed in-context learning algorithm relies on LLM's inference capabilities. It avoids the complexity of tedious model training and hyper-parameter fine-tuning, which is a well-known bottleneck of many ML algorithms.
Specifically, the proposed algorithm first describes the target task via formatted natural language, and then designs the in-context learning framework and demonstration examples. 
After that, it considers two cases, namely discrete-state and continuous-state problems, and proposes state-based and ranking-based methods to select appropriate examples for these two cases, respectively.   
Finally, the simulations demonstrate that the proposed algorithm can achieve satisfactory performance without \blue{updating the LLM model parameters.}
%
Such an efficient and low-complexity approach has great potential for future wireless network optimization.
\end{abstract}


\section{Introduction}

The envisioned 6G network will be increasingly complicated with diverse application scenarios and novel signal processing techniques, e.g., vehicle-to-everything (V2X), mmWave and THz networks, reconfigurable intelligent surface, etc \cite{zhang20196g}.   
The constantly evolving network architecture requires more efficient management schemes, and most existing network optimization methods can be summarized into two main approaches: convex optimization and machine learning (ML) algorithms.
Specifically, convex optimization usually needs dedicated problem formulation for each specific task, then transforms the objective function or constraints into convex forms. 
By contrast, ML algorithms, such as reinforcement learning, have lower requirements for problem formulations, but the tedious model training and fine-tuning indicate a large number of iterations \cite{zhou2023survey}.
Therefore, these potential issues, e.g., problem-specific transformation and relaxation, hyperparameter tuning, and long training iterations, have become obstacles to further improve the efficiency of next-generation networks.

Recently, generative AI (GAI) and large language models (LLMs) have provided promising opportunities for network fields \cite{zhou2024large}, e.g., 6G edge intelligence\cite{lin2023pushing}, beamforming\cite{zhang2024generative}, reconfigurable intelligent surfaces (RISs)\cite{chaaya2024ris}, wireless network design \cite{qiu2024large}, etc. 
Motivated by the issues of existing optimization techniques, this work explores LLM-enabled network optimization techniques.
It considers base station (BS) power control as a case study, which is a fundamental and critical technique that has been extensively studied by using convex optimization, game theory, reinforcement learning, etc. 
However, few existing studies have addressed this crucial network optimization problem from a language-related perspective. 
Such a novel technique has great potential to save human labour for network operations, e.g., i.e., optimizing network performance by using natural language directly. LLMs can also provide detailed explanations for their outputs, helping humans understand complicated 6G networks.

To this end, this work proposes a novel LLM-enabled in-context learning algorithm for optimization tasks.  
In-context learning indicates learning from language-based descriptions and demonstrations, which has multiple advantages \cite{dong2022survey}: 
1) In-context learning relies on LLM's inference process, and it avoids the complexity of \blue{updating the LLM model parameters, saving considerable computational resources}; 
2) In-context learning allows natural language-based task design and implementation, and the operator can easily formulate the target task using human language and instructions.
%
In addition, prompt engineering only requires forward passing of the model without the need for backpropagation. Therefore, the fast implementation and low response time can more efficiently handle network dynamics. 

In particular, our proposed technique first designs a natural language-based task description, i.e., task goal, definition, and rules. 
The formatted task description, along with a set of selected examples, will become the prompt input for the LLM model.
Then, the LLM model can utilize the task description and advisable examples to generate a decision based on the current environment state.
Different from existing LLM-enabled network optimization studies\cite{qiu2024large}, we propose a novel experience pool framework. It will collect the previous experience and decisions of LLMs, serving as references for future decision-making.
In addition, examples are crucial for in-context learning. Distinct from prior studies\cite{zhang2024generative, qiu2024large}, we further propose two novel example selection methods, namely state-based and ranking-based approaches, for discrete-state and continuous-state problems, respectively. 
With experience pools and proper example selection, LLMs can utilize the accumulated experience and find hidden patterns from examples, making optimal decisions accordingly.

The core contribution of this work is that we proposed a LLM-enabled 
in-context learning technique for network optimization, which can learn from the language-based task descriptions and environment interactions. 
It overcomes the tedious model training and parameter fine-tuning processes, which are usually time-consuming in conventional ML algorithms.
We further evaluate the proposed algorithm with various LLMs, e.g.,  Llama3-8b-instruct, Llama3-70b-instruct, and GPT-3.5 turbo, and the simulations prove that the proposed algorithm can achieve satisfactory performance.

\section{System Model}

\subsection{Power Control Problem Formulation}
This section introduces a BS power minimization problem, serving as a case study of the proposed in-context learning algorithm.
Considering a BS with $U_b$ users, the achievable data rate $C_{b,u}$ between BS $b$ and user $u$ is defined by
\begin{equation}
\label{eq4}
\resizebox{0.91\hsize}{!}{$
C_{b,u}=\sum\limits_{k=1}\limits^{K_{b}}d_{k}log(1+ \frac{p_{b,k}h_{b,k,u}\gamma_{b,k,u}}{\sum\limits_{b'\in B_{-b}}{p_{b',k'}h_{b',k',u'}\gamma_{b',k',u'}}+d_{k}N_{0}}),$}
\end{equation}
where $K_{b}$ is the total number of resource blocks (RBs) in BS $b$, $d_{k}$ is the bandwidth of RB $k$, 
$p_{b,k}$ indicates the transmission power of BS $b$ on RB $k$, 
$h_{b,k,u}$ defines the channel gain between BS $b$ and user $u$ on RB $k$, and $N_{0}$ is the noise power density. 
For the RB allocation, $\gamma_{b,k,u} \in \{0,1\}$ indicates whether RB $k$ is allocated to the transmission for user $u$.  
For the interference, $B_{-b}$ represent the set of adjacent BSs except for BS $b$, $p_{b',k'}h_{b',k',u'}\gamma_{b',k',u'}$ defines the inter-cell interference, and we assume orthogonal frequency-division multiplexing is applied to eliminate intra-cell interference.

This work aims to minimize the BS transmission power and meanwhile satisfy the average data rate constraint \cite{chiang2008power}:
\begin{subequations}\label{e5:main}
\begin{align}
\min_{P_{b}} & \enspace \sum\nolimits_{b\in B} P_{b}     & \tag{\ref{e5:main}} \\
\text{s.t.}  & \enspace 0 \leq P_{b} \leq P_{max},  & \label{e5:a}\\
& \enspace  P_{b}=\sum\nolimits_{k=1}^{K_{b}} p_{b,k},  & \label{e5:b}\\ 
& \enspace \sum\nolimits_{u=1}^{U_{b}}C_{b,u}/U_{b}  \geq C_{min}, & \label{e5:c}   
\end{align}
\end{subequations}
where $P_b$ is the total transmission power of BS $b$ and $P_{b}=\sum\nolimits_{k=1}^{K_{b}} p_{b,k}$, $p_{b,k}$ has been defined in equation (\ref{eq4}) as the transmission power of RB $k$, 
$P_{max}$ is the maximum power, 
$U_b$ is the total number of users, and $C_{min}$ is the average achievable data rate constraint.
We assume $P_b$ is equally allocated to all RBs, and a proportional fairness method is used for RB allocation, which has been widely used as a classic approach. Then we can better focus on LLM features. 

%

\subsection{Language-based Power Control Task Description}
\label{sec-task}

Problem (\ref{e5:main}) has been extensively investigated in existing studies, but this work differs from previous works by presenting a unique view from the perspective of LLM-enabled network optimization. 
Instead of defining specific equations as in (\ref{e5:main}), here we use natural language to describe the optimization task. Specifically, the defined task description is shown below, which will further be used to prompt LLMs:
\begin{tcolorbox}[title = {Task description for BS transmission power control}]
\label{box1}
\textbf{Task goal}: You have a decision-making task for base station power control, and you need to select between 4 power levels from 1 to 4.\\
\textbf{Task definition}: You have to consider the specific user number of each case, which is the “BS user number”.\\
Following are some examples $\{Example\_set\}$.\\
Now I will give you a new condition to solve, the current BS user number is $\{Num\_BS\_user\}$. \\
\textbf{Rules}: Now please select from “level 1”, “level 2”, “level 3”, and “level 4” based on the above examples.
\end{tcolorbox}

In particular, the $Task\_goal$ first specifies a “\textit{decision-making task for base station power control}”, and the goal is to “\textit{select between 4 power levels}”.
Then the $Task\_definition$ introduces the environment states we need to consider. For example, this work assumes the total user numbers may change dynamically, and then the LLM has to consider the “\textit{user number}” of each case. 
%
After that, the example set $\mathcal{E}_{t}$ is included by “\textit{Following are some examples....}”, and we provide a new condition for the LLM to solve with the current user number $U_b$. 
Finally, we set extra reply rules such as “\textit{select from ... based on the above examples}”, indicating the LLM to focus on the decision-making process.   
Such a definition provides a standard template for addressing many optimization tasks by including goals, definitions, and rules.

\begin{figure*}[t]
\centering
\setlength{\abovecaptionskip}{4pt} 
\includegraphics[width=0.95\linewidth]{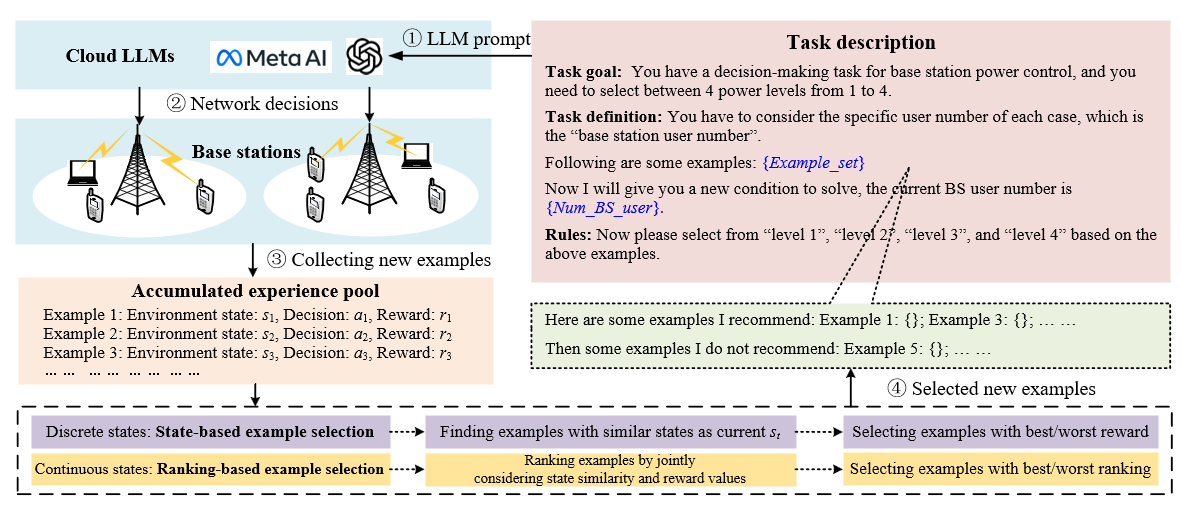}
\caption{Overall design of the proposed LLM-enabled in-context learning for transmission power control.}
\label{fig-optimization}
\vspace{-15pt}
\end{figure*}

\section{In-context Learning-based Optimization Algorithm}
\label{sec-conext}


\subsection{In-context Learning}
\label{sec-incontext}
In-context learning refers to the process that LLMs can learn from formatted natural language such as task descriptions and task solution demonstrations, to improve the performance on target tasks.
In-context learning can be defined as \cite{dong2022survey}
\begin{equation}
\label{eq_llm}
D_{task} \times \mathcal{E}_{t} \times s_{t} \times \mathcal{LLM} \Rightarrow a_{t},
\end{equation}
where $D_{task}$ is the task description and query, $\mathcal{E}_{t}$ is the set of examples at time $t$, $s_{t}$ is the environment state at time $t$ that is associated with the target task,
$\mathcal{LLM}$ indicates the LLM model, and $a_{t}$ is the LLM output. 
Here we expect the LLM can utilize the initial task description $D_{task}$, learn from the example set  $\mathcal{E}_{t}$, and then make decision $a_{t}$ based on current environment state $s_{t}$ of the target task.

The LLM's in-context learning capabilities can be explained by implicit fine-tuning according to \cite{dai2022can}. Specifically, LLMs will produce meta-gradients based on given examples $\mathcal{E}$ by forward computation, and then the meta-gradients are applied by using the attention mechanism to build an in-context learning model \cite{dai2022can}:
\begin{equation}\label{eq-linear}
    \tilde{f}_{\text{ICL}}(\mathbf{q})= \mathbf{q} \left( W_{\text{ZSL}} + \Delta W_{\text{ICL}} \right),
\end{equation}
where $\textbf{q}$ is the query vector in the attention mechanism, $W_{ZSL}$ indicates the zero-shot learning case without examples, $\Delta W_{\text{ICL}}$ is the updated weight when examples $E \in \mathcal{E}$ are provided by in-context learning.

\subsection{Examples and Optimization Framework Design}
The analyses in Section \ref{sec-incontext} show that examples are of great importance in in-context learning, which will directly affect the $\Delta W_{\text{ICL}}$ values. 
Here we define an example by
\begin{equation}
\label{eq_example}
E=\{s, a, r(s,a)\}, E \in \mathcal{E}    
\end{equation}
where $s$ and $a$ are environment state and decision, respectively. Inspired by reinforcement learning, we further define a reward value to evaluate the decision $a$ by
\begin{equation}
\label{eq_reward}
r=P_{target}-P_{b}-\beta   
\end{equation}
where $P_{target}$ is a target power consumption, and $P_b$ has been defined in problem (\ref{e5:main}) as the total power consumption of BS $b$. $\beta$ is a penalty term, which is only applied when constraint  (\ref{e5:c}) is not satisfied. Then, $r$ provides a comprehensive metric to evaluate the selected decision $a$ under environment state $s$.

Fig.\ref{fig-optimization} shows the overall design of the proposed in-context learning algorithm. 
Specifically, the above task description $D_{task}$, current environment state $s_t$, and selected examples $\mathcal{E}_{t}$ are integrated as input prompt as defined in equation (\ref{eq_llm}), and then the LLM model will generate a power control decision $a_t$ based on $s_t$ and the experiences in $\mathcal{E}_t$.
Then, the decision $a_t$ is implemented, the achieved data rate $C_{b,u}$ is collected, and the reward $r_t$ is calculated as equation (\ref{eq_reward}). $E_{t}=\{s_t, a_t, r_{t}(s_t,a_t)\}$ becomes a new example in the accumulated experience pool $\mathcal{E}_{pool}$. 
%
After that, 
we considered two scenarios, namely discrete and continuous state problems, and proposed state-based and ranking-based example selection methods.
Based on the next environment state $s_{t+1}$, a new example set $\mathcal{E}_{t+1}$ is selected, and the selected examples are inserted into the task description with $s_{t+1}$, becoming a new prompt for the LLM model to generate $a_{t+1}$. 


\subsection{State-based Example Selection for Discrete States}
\label{sec-discrete}
Selecting appropriate examples is critical for in-context learning. To improve the quality of selected examples, this subsection introduces a state-based example selection method for problems with discrete environment states.
Considering a target task with environment state value $s_{target}$, the set of relevant examples can be identified by
\begin{equation}
\label{eq_example_re}
\mathcal{E}_{relevant}= \Big \{E\{s,a,r(s,a)\} \Big| s=s_{target}, E \in \mathcal{E}_{pool} \Big\}
\end{equation}
where $\mathcal{E}_{pool}$ is the accumulated experience pool in Fig. \ref{fig_all}. 
Given the current state $s_{target}$, equation (\ref{eq_example_re}) provides a practical solution to find the most relevant examples $\mathcal{E}_{relevant}$.
With $\mathcal{E}_{relevant}$, we can easily select recommended top examples with higher reward, and inadvisable examples with lower reward or violating the minimum data rate constraint in the problem formulation.  
In addition, we include a well-known epsilon-greedy policy to balance exploration and exploitation. 
\begin{equation} \label{eq_epsilon}
a=\left\{
\begin{array}{lcl} \text{Random action selection},     & \text{if }  rand < \epsilon; \\
\text{LLM-based decision-making,} &  \text{else},
\end{array} \right.
\end{equation}
where $\epsilon$ is a predefined value, and $rand$ is a random number between 0 and 1. 
Therefore, the random exploration in equation (\ref{eq_epsilon}) can constantly explore new examples, and then the LLM model can learn from better relevant examples $\mathcal{E}_{relevant}$ to improve the performance.

\subsection{Ranking-based Example Selection for Continuous States}
\label{sec-conti}
This subsection introduces a ranking-based method to select proper examples for continuous state problems. Specifically, continuous states indicate an infinite number of possible examples, and identifying the most relevant and high-quality examples can be challenging.
For instance, when using average user-BS distance as an environment state for BS transmission power control with a target task $s_{target}$, it is unlikely to find a specific existing example $E\{s,a,r(s,a)\}$ with $s=s_{target}$, since $s_{target}$ is a random number within the BS maximum coverage distance. 
%
To this end, we define a new metric $\mathcal{L}$ for example selection with continuous states:
\begin{equation} \label{eq-metric}
\mathcal{L}(E,s_{target})=r(s,a)-\tau||s-s_{target}||,    
\end{equation}
where $\mathcal{L}(E,s_{target})$ is a comprehensive metric to evaluate the usefulness of $E=\{s, a, r(s,a)\}$ to the decision-making of $s_{target}$, 
and $||s-s_{target}||$ is the $L^2$ norm to define the distance between $s$ and $s_{target}$. 
Equation (\ref{eq-metric}) aims to jointly consider the reward and states of example $E$, and $\tau$ is a weighting factor to balance the importance of higher reward $r(s,a)$ and more similar states between $s$ and $s_{target}$. 
Specifically, a higher reward $r(s,a)$ indicates that $E$ includes a good action selection $a$ under environment state $s$, 
and meanwhile lower $||s-s_{target}||$ value means the environment state $s$ in $E$ is more similar to $s_{target}$.
Therefore, $\mathcal{L}(E, s_{target})$ becomes a comprehensive metric to evaluate the quality of examples in $\mathcal{E}_{pool}$. With $\mathcal{L}(E, s_{target})$, we can easily select recommended and inadvisable examples by ranking all elements in $\mathcal{E}_{pool}$.


\begin{figure}[t]
\centering
\includegraphics[width=0.9\linewidth]{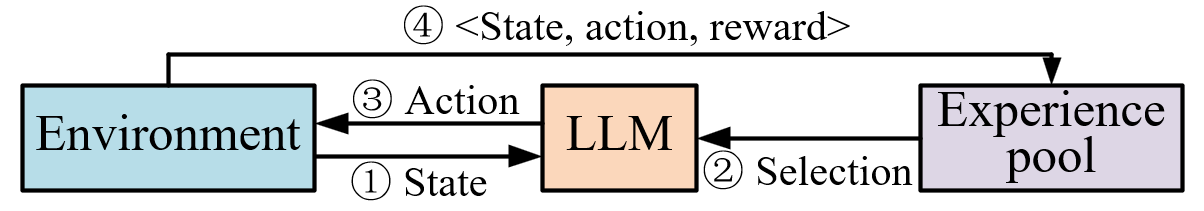}
\caption{\blue{\textbf{The overall procedure of the example-related scheme.}}}
\label{fig-example}
\vspace{-15pt}
\end{figure}

\subsection{\blue{Computational Complexity Analyses}}

\blue{Fig. \ref{fig-example} summarizes the overall procedure of example-related schemes. In particular, the LLM receives the state from the environment, and then uses the examples provided by the experience pool to select actions such as the transmission power level. The implementation results will become a new example for the pool. 
Meanwhile, no additional computational cost is incurred for example selection, as each new example is simply appended to the accumulated experience pool after implementation. 
Secondly, for example selection in discrete state problems, it is easy to search the experience pool to identify $s=s_{target}$. For continuous states, we calculate the $\mathcal{L}(E,s_{target})$ metric for all examples in the pool, and then select the best examples accordingly. 
Therefore, the cost of example selection follows a linear complexity.
Finally, note that the LLM inference time is affected by model architecture, hardware constraints, and task types, and it can also be further optimized by quantization, sparsity exploitation, and architectural innovations. }

\section{Performance Evaluation}

\begin{figure*}[!t]
\centering
\subfigure[Discrete state space: System reward and \newline service quality comparison.]{
\includegraphics[width=5.5cm,height=4.1cm]{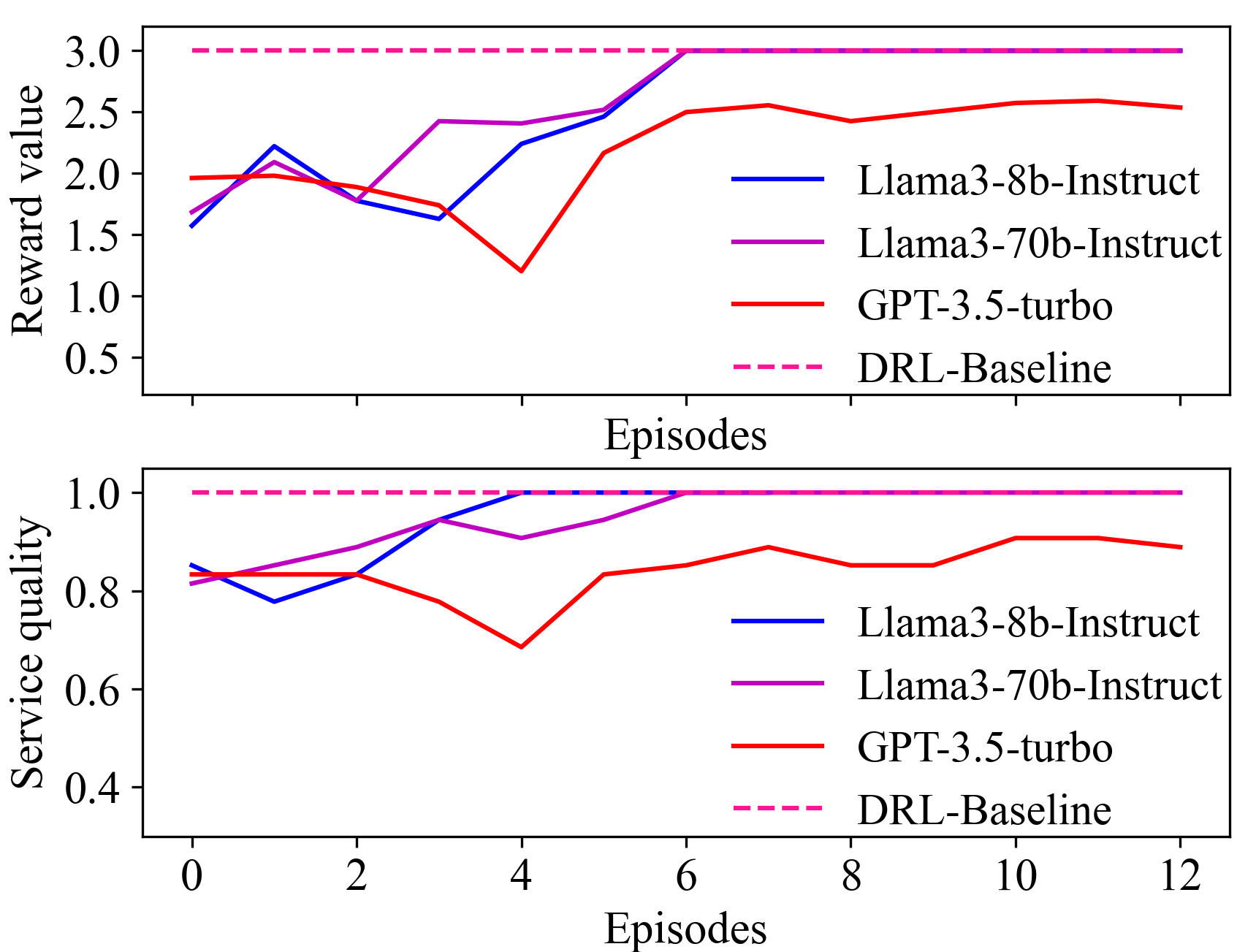} 
\label{f-r1}
}
\centering
\subfigure[\blue{{Continuous state space: System reward and \newline service quality comparison.}}]{
\includegraphics[width=5.5cm,height=4.1cm]{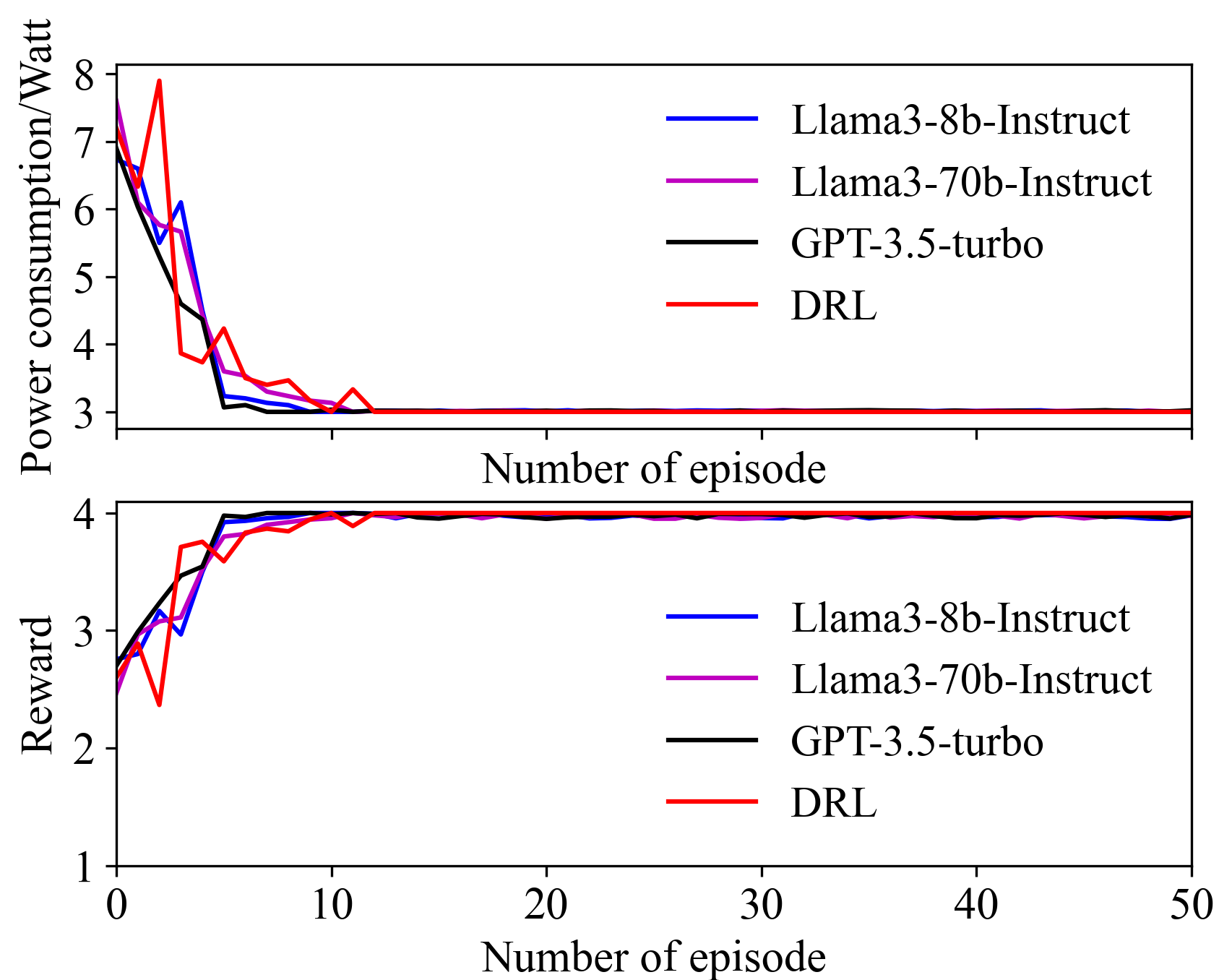} \label{f-r2}
}
\subfigure[{\blue{{Continuous state space: Average reward comparison under different data rate constraints. (ICL: In-context learning)}}}]{
\includegraphics[width=5.5cm,height=4.1cm]{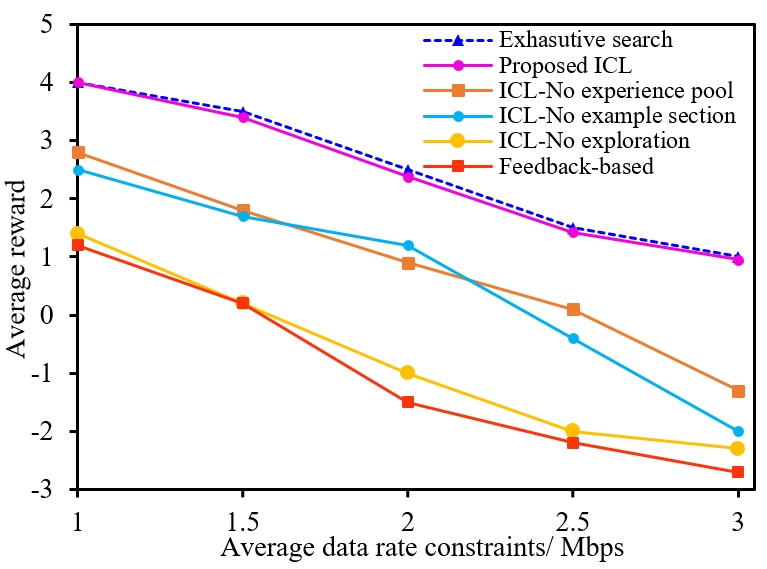} \label{f-r3}
}
\quad
\subfigure[\blue{{Continuous state space: Average power \newline consumption comparison under different \newline data rate constraints.}}]{
\includegraphics[width=5.5cm,height=4.1cm]{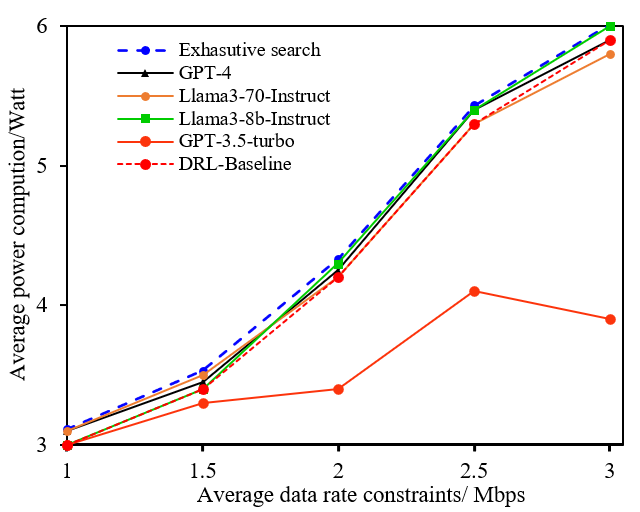} \label{f-r5}
}
\subfigure[\blue{{Continuous state space: Average service quality comparison under different data rate constraints.}}]{
\includegraphics[width=5.5cm,height=4.1cm]{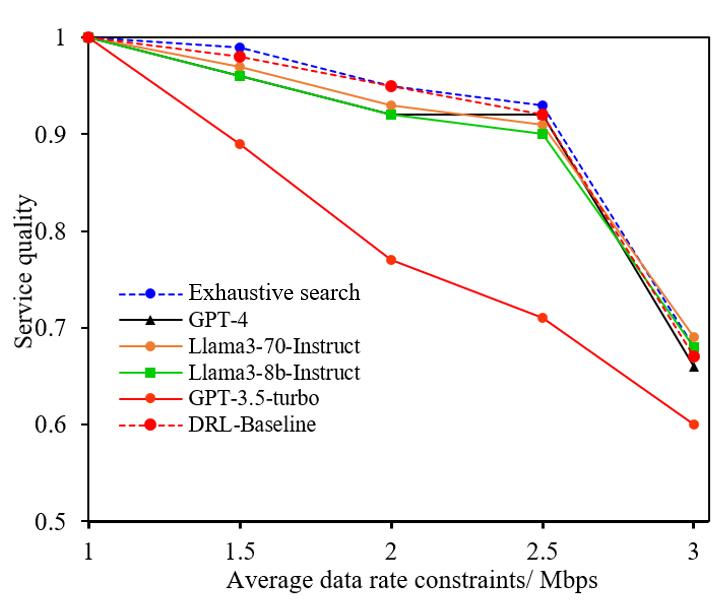} \label{f-r6}
}
\subfigure[\blue{{Continuous state space: Reward with increasing number of examples and and larger spaces.}}]{
\includegraphics[width=5.5cm,height=4.1cm]{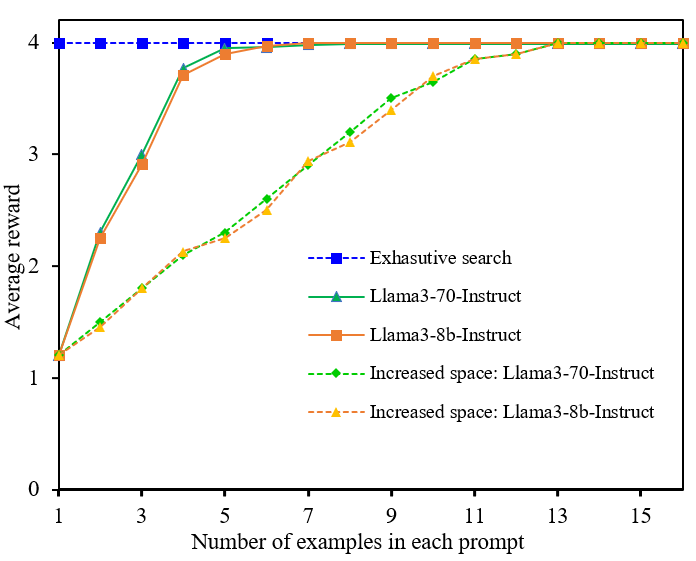} \label{f-r4}
}
\caption{Simulation results and comparisons}
\label{fig_all}
\vspace{-13pt}
\end{figure*}

\subsection{Simulation Settings}
We consider three adjacent small base stations (SBSs), the user number of each SBS randomly changes from 5 to 15, and the SBS's coverage is 20 meters. The channel gain applies 3GPP urban network models, and
2 cases are evaluated: \\
\textbf{Case I}: Discrete states defined by user numbers of each SBS;\\ 
\textbf{Case II}: Continuous states defined by average user-SBS distance, which represents 2 kinds of network optimization problems.
Then, the simulation considers 3 main approaches: \\ 
\blue{\textbf{1) LLM-based method} applies our proposed technique with various models: Llama3-8b-instruct, Llama3-70b-instruct, GPT-4, and GPT-3.5 turbo. Using LLM models with various sizes and capabilities can better evaluate the performance of our proposed algorithms. 
We have also evaluated the system performance by ablation studies, e.g., performance without the proposed mechanisms such as experience pool, example selection, and random exploration. In addition, we considered the feedback-based approach in \cite{qiu2024large} as another baseline. } 
\textbf{2) DRL-based method}: We employ DRL as a baseline algorithm, since it has been widely used to address various network optimization problems in many existing studies \cite{zhou2023survey}.
\blue{\textbf{3) Exhaustive search}: We apply exhaustive search method as the optimal baseline, searching for the best decisions exhaustively.}

\subsection{Simulation Results}
Fig. \ref{fig_all} shows the simulation results and comparisons.
\blue{Firstly, Fig. \ref{f-r1} and \ref{f-r2} present the reward and service quality under discrete and continuous state spaces.} 
\blue{One can observe that LLMs can achieve higher rewards as the number of episodes increases, and Llama3 LLMs present close performance as the DRL baseline method for discrete and continuous problems.}
Fig. \ref{f-r1} and \ref{f-r2} demonstrate that LLMs can learn from previous examples and interactions, and then improve their performance on target tasks iteratively.

\blue{Then, we implement ablation studies in Fig. \ref{f-r3}. 
It demonstrates the importance of our proposed techniques, e.g., the experience pool design, the example selection strategies, and exploration policies. 
Without these designs, the in-context learning technique presents a much lower reward than exhaustive search. 
It highlights the necessity of our designs in understanding the internal mechanisms of in-context learning technique and LLM-enabled optimization.
Meanwhile, the feedback-based method also shows a worse performance. It means that using the feedback from previous implementations solely cannot fully reflect the complexity of a dynamic environment.   
It can be used to address static optimization problems as introduced in \cite{qiu2024large}, but it cannot handle dynamic optimization problems as defined in our work.}

Moreover, we observe the algorithm performance under different minimum data rate constraints. Fig. \ref{f-r5} and \ref{f-r6} present the average power consumption and service quality, respectively. 
%
As expected, given the limited bandwidth, increasing the minimum data rate constraint leads to higher power consumption and
lower service quality for all algorithms.
\blue{GPT-4, Llama3-8b, and Llama3-70b show a close performance as the task-specific DRL algorithm and exhaustive search baseline.}
They demonstrate that the proposed in-context learning can adapt to different optimization settings and then adjust their policies to improve the performance of target tasks. 
By contrast, GPT-3.5 shows worse performance than other techniques, which indicates that algorithm performance is also related to specific LLMs. Specifically, \blue{GPT-4 and Llama3} represent state-of-the-art LLM designs, while GPT-3.5 is an early and outdated LLM model.

\blue{Finally, Fig. \ref{f-r4} evaluated the system performance with enlarged state space and changing number of examples in the prompt. Firstly, one can observe that increasing the number of examples can constantly improve the average reward. However, such improvement becomes less obvious when plenty of examples are provided.
On the other hand, increasing the state space means that more examples are needed in the prompt to achieve a satisfactory performance, e.g., more references and experience are needed to make proper decisions. 
However, it is worth noting that the overall performance is still constantly improving by increasing the number of provided examples, and it finally achieves a comparable performance as the exhaustive search method. 
}




\section{Conclusion}
LLM is a promising technique for future wireless networks, and this work proposes an LLM-enabled in-context learning algorithm for BS transmission power control.
The proposed algorithm can handle both discrete and continuous state problems, and the simulations show that it achieves comparable performance as conventional DRL algorithms. This work demonstrates the great potential of in-context learning for handling network management and optimization problems. 
In the future, we will focus on the practical application of LLMs to wireless networks, including operation costs, on-premises deployment, and real-time performances.

\normalem
\bibliographystyle{IEEEtran}
\bibliography{Reference}

\end{document}